# Influence of target thickness on the release of radioactive atoms

Julien Guillot[a*], Brigitte Roussière[a], Sandrine Tusseau-Nenez[b], Nicole Barré-Boscher[a], Elie Borg[a] and Julien Martin[a]

[a] Institut de Physique Nucléaire CNRS/IN2P3 UMR 8608 – Université Paris Sud – Université Paris Saclay, F-91406 Orsay Cedex France
[b] Physique de la Matière Condensée Ecole Polytechnique/CNRS UMR 7643 – Université Paris Saclay, F-91128 Palaiseau Cedex France

*Corresponding author: guillotjulien@ipno.in2p3.fr

___________________________________________________________________________

**Abstract**

Nowadays, intense exotic beams are needed in order to study nuclei with very short half-life. To increase the release efficiency of the fission products, all the target characteristics involved must be improved (e.g. chemical composition, dimensions, physicochemical properties such as grain size, porosity, density…). In this article, we study the impact of the target thickness. Released fractions measured from graphite and uranium carbide pellets are presented as well as Monte-Carlo simulations of the Brownian motion.

___________________________________________________________________________

___________________________________________________________________________

At the ALTO (Accélérateur Linéaire et Tandem d'Orsay) facility of IPNO (Institut de Physique Nucléaire d'Orsay), the production of radioactive beams is achieved by the ISOL method (Isotope Separation On Line). An incident beam impinges on so-called thick targets to produce the maximum number of radioactive atoms by nuclear reactions (fragmentation, spallation and fission) [1,2]. The intensity of the obtained radioactive beams is defined by:

$I = I_P \cdot \sigma \cdot N \cdot \varepsilon_r \cdot \varepsilon_{ion} \cdot \varepsilon_{tr}$, where $I_P$ is the intensity of the incident particle beam delivered by the accelerator, $\sigma$ the cross section of the isotopes of interest, $\varepsilon_r$ the release efficiency from the target to the ion source for the element of interest, $\varepsilon_{ion}$ the ionization efficiency of this element and $\varepsilon_{tr}$ the transport efficiency of the separator.

The major factor limiting the intensity of the beam is the release efficiency, especially for short-lived isotopes. Indeed, when a radioactive atom is produced in the target, it must be released before decaying.

At ALTO, thick targets are composed mainly of uranium carbide (noted $UC_x$ as $UC_2$ is stabilized with UC in minor quantity) obtained by reacting uranium dioxide and carbon. The synthesis reaction of carburization is given by:

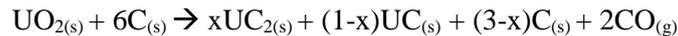

$UO_{2(s)} + 6C_{(s)} \rightarrow xUC_{2(s)} + (1-x)UC_{(s)} + (3-x)C_{(s)} + 2CO_{(g)}$

The main factors influencing the release efficiency are the grain size of $UC_x$, the homogeneity of the microstructure, the open pore size distribution obtained during the degassing [3,4].

In this paper, the effect of the target thickness on the release of fission products from $UC_x$ target is discussed. Since the $UC_x$ targets used have graphite in excess, diffusion in graphite can give an idea of the diffusion in the target. Therefore the europium release from Eu-doped graphite pellets with different thicknesses was also studied and the released fractions (RFs) measured were compared to the Monte-Carlo calculation results of the Brownian motion.

Four identical mixtures were prepared, each containing 10 g of graphite powder (Cerac, purity = 99.5%, 325 mesh) and 50 mg of $Eu_2O_3$ powder. Among the lanthanides (Ln), europium was chosen because it is one of the most volatile [5] and we had already performed experiments on Ln-RFs from graphite pellets [6]. The powders were mixed for 5 minutes under ultrasound with 40 mL of isopropanol then dried over a sand bath. Each mixture was pressed in a 13 mm-diameter mold using a manual press at 220 MPa for 12 seconds. Four sets of pellets with mean thickness of 4 mm, 2 mm, 1 mm and 0.7 mm were thus obtained. Several tests were performed at different temperatures and heating times in order to determine the values leading to a measurable evaporation of europium by the four types of pellets. Finally, each series of pellets was heated under high vacuum ($10^{-6}$ mbar) at 1200 °C for 4 hours. All the pellets were weighed before and after heating using a precision balance (Mettler Toledo XP6) to determine the percentage of europium released.

Table 1 summarizes for the four batches of pellets the number of studied pellets, their average thickness and the deduced RFs. The main error sources in determining RFs are due to the temperature gradient inside the oven and to the pellet-thickness dispersion. Therefore the higher the pellet number used for the measurement, the greater the RF error bar. But the RF variation calculated as a function of the sample thickness is not linear (see the simulation results below). Taking into account all these aspects the RF error bars given in table 1 were estimated. We can notice that the europium RFs increases when the pellet thickness decreases.

Table 1. Characteristics of the graphite pellets.

| Number of pellets | Thickness(mm) | RF of Eu (%) |
|---|---|---|
| 42 | 0.678 ± 0.094 | 87.7 ± 10.0 |
| 20 | 1.153 ± 0.103 | 57.1 ± 17.0 |
| 10 | 2.180 ± 0.144 | 47.6 ± 8.3 |
| 5 | 4.286 ± 0.160 | 31.8 ± 5.6 |

To calculate the RF as a function of the thickness of the target, the europium particle path in the graphite pellet was described by a Brownian motion using Monte-Carlo methods. The pellets with different thicknesses were modeled as cylinders. Each simulation consisted in 10000 trials. In each case, a particle was created at an initial position chosen randomly inside the cylinder. As long as the particle is far from the pellet boundaries, Brownian motion is replaced by an equivalent "spherical process" in order to cross the medium within a reasonable computation time [7]. Then the particle exits the pellet doing small random moves. The number of steps performed before exiting the cylinder was recorded. After simulations with different cylinder sizes, the RF curves can be drawn as a function of the step number (Figure 1). The RF values obtained experimentally are also reported. The four experimental points are expected to be aligned with a given step number, indeed in such approach, a given temperature maintained over a specified time will allow the particle to execute a well-defined step number. The experimental values correspond to step number close to $6.45 \times 10^{11}$ (the vertical dotted line).

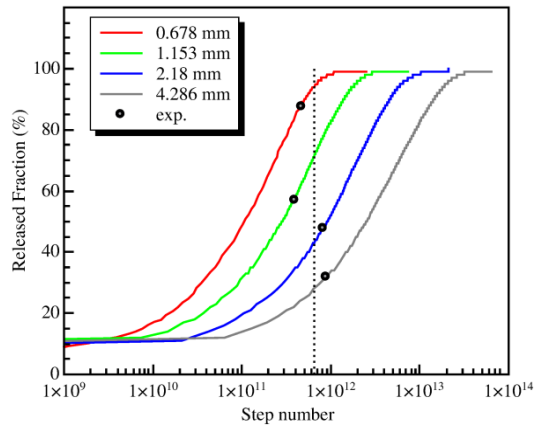

Figure 1: Simulation by a Brownian motion of the RF as a function of the step number.

A second simulation was performed by setting the maximum step number to $6.45 \times 10^{11}$ and by varying the thickness of the cylinder. The principle was the same as before and consisted in creating randomly 10000 particles in a cylinder and moving them randomly. The RF is given by the ratio between the number of particles that exited the cylinder within the $6.45 \times 10^{11}$ steps over the total trial number. The results of this simulation are plotted in Figure 2, as well as the experimental RFs.

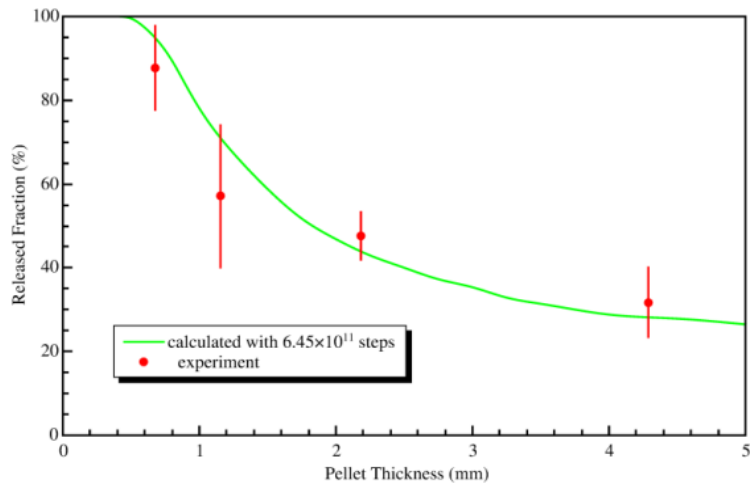

Figure 2: Simulation by a Brownian motion of the RF as function of the pellet thickness.

The calculations are in agreement with the experimental data. These results indicate that the assumption of isotropic diffusion in homogeneous medium well describes the europium diffusion in graphite.

The subsequent topic evaluated in this work was the effect of the target thickness on the release of fission products from $UC_x$ targets. The $UC_x$ pellets were prepared by adapting a protocol developed at CERN ISOLDE on lanthanum carbide. Uranium dioxide powder (AREVA sample No. 65496) used contains 0.25 wt% of $^{235}U$ and impurities of Cr, Ni and Fe respectively at 4, 6 and 16 µg/gU. This powder was dry milled for 4 hours following the protocol described in ref. [8]. The milled powder was dispersed in isopropanol under ultrasound and then mixed with carbon nanotubes (Nanocyl, purity > 95%, 10 nm - size 1.5 µm) in C/U molar ratio equal to 5, 6 or 7. Before planetary mixing, carbon nanotubes were dispersed under agitation and ultrasound in isopropanol. This mixture was then dried using a distillation column. The obtained powder was collected and pressed into pellets at 220 MPa for 12 seconds in a 13 mm diameter mold using a semi-automatic press (SPECAC Automatic Hydraulic Press). The pellets were sintered under vacuum ($10^{-6}$ mbar) at 1800 °C for 2 hours. The pellets were characterized by their geometrical dimensions (diameter, thickness) and weight. They were analyzed with nitrogen gas adsorption and desorption using the Brunauer Emmett Teller method (BET method - Micromeritics ASAP2020) and helium pycnometry (Micromeritics AccuPyc II 1340) before and after heating, to determine respectively the specific surface area (SSA) of the sample and its corresponding apparent grain size, and its open and close porosities [9] . Finally the pellets were irradiated 20 minutes by fast neutrons produced in the break up of 26 MeV deuterons delivered by the IPNO Tandem and the rate of fission products released by the samples after heating at 1700 °C during 30 minutes was determined by γ-spectroscopy measurements. The experimental conditions and analysis procedure were previously detailed [4].

The physicochemical characteristics measured after sintering for the three samples designated by their initial C/U molar ratio are summarized in Table 2. The grain size, the porosity and the mass of the three samples are very similar but the C/U molar ratios are different, which implies that the pellet dimensions are different.

Table 2. Characteristics of the $UC_x$ pellets.

| Sample | Mass (g) | Pellet size | | BET | | He Pycnometry | |
|---|---|---|---|---|---|---|---|
| | | Diameter (mm) | Thickness (mm) | SSA ± 5% (m²/g) | Grain size (nm) | Open ± 3% (%) | Closed ± 3% (%) |
| C/U = 5 | 0.3268 | 11.03 | 1.23 | 11.53 | 139.69 | 64 | 8 |
| C/U = 6 | 0.3270 | 12.37 | 1.62 | 19.56 | 139.21 | 68 | 12 |
| C/U = 7 | 0.3263 | 13.01 | 1.94 | 31.87 | 131.90 | 68 | 15 |

The RFs obtained for fifteen elements are shown in Figure 3. Taking into account that fission products diffuse more quickly in graphite than in uranium, the RFs are expected to increase when the amount of graphite in the sample (and thus the C/U molar ratio) increases. On the other hand, the study of diffusion in graphite showed that the RF decreases when the pellet thickness increases. One can notice that, without correction from the pellet dimensions (Figure 3a), there is a compensation between the effect of the C/U molar ratio and the pellet thickness. Therefore the RFs overlap. When the RFs are corrected from pellet dimensions (diameter and thickness set respectively as 13 mm and 1.7 mm) using the model described above, significant differences appear (Figure 3b), and then the effect of the C/U molar ratio is clearly recovered.

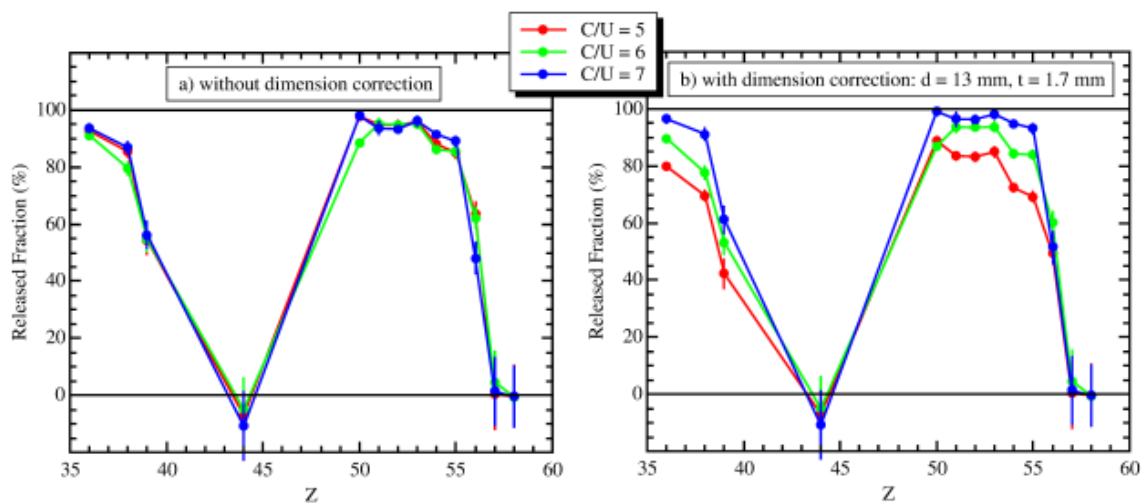

Figure 3: Release of fission products from $UC_x$ targets a) without and b) with dimension correction.

This study shows that the thickness of the target seems to have a real impact on the release efficiency. In this respect worldwide facilities such as SPES (Selective Production of Exotic Species) at INFN-LNL (Istituto Nazionale di Fisica Nucleare - Laboratori Nazionali di Legnaro) have specifically chosen thin targets to enhance isotope release [10]. In order to obtain intense radioactive ion beams of short-lived fission products, the $UC_x$ pellets will need to be as thin as possible while being easy to handle and must contain an excess of carbon. But it remains to determine to what extent the porosity may counteract the effect of the pellet thickness on the released fractions.

## Acknowledgements

We thank the NESTER, SPR and ALTO teams for advises and discussions on this project and especially Abdelhakim Saïd, Thony Corbin, Christophe Planat, Robert Leplat, Christophe Vogel, Alain Semsoun, Sébastien Wurth and David Verney for their help during the experiments. This work was supported by IPNO.